\documentclass[12pt]{article}
\usepackage{graphicx}
\usepackage{hyperref}

\begin{document}

\title{When will two agents agree on a quantum measurement outcome?
  Intersubjective agreement in QBism}

\author{R\"udiger Schack\thanks{Email: r.schack@rhul.ac.uk} \\
{\it Department of Mathematics, Royal Holloway, University of London} \\
{\it Egham, Surrey TW20 0EX, UK}}

\date{12 December 2023}
\maketitle

\begin{abstract}
In the QBist approach to quantum mechanics, a measurement is an action an agent
takes on the world external to herself. A measurement device is an extension of
the agent and both measurement outcomes and their probabilities are personal to
the agent. According to QBism, nothing in the quantum formalism implies either
that the quantum state assignments of two agents or their respective
measurement outcomes need to be mutually consistent. Recently, Khrennikov has
claimed that QBism's personalist theory of quantum measurement is invalidated
by Ozawa's so-called intersubjectivity theorem. Here, following Stacey, we
refute Khrennikov's claim by showing that it is not Ozawa's mathematical
theorem but an additional assumption made by Khrennikov that QBism is
incompatible with. We then address the question of intersubjective agreement in
QBism more generally. Even though there is never a necessity
for two agents to agree on their respective measurement outcomes, a QBist agent
can strive to create conditions under which she would expect another agent's
reported measurement outcome to agree with hers. It turns out that the
assumptions of Ozawa's theorem provide an example for just such a condition.
\end{abstract}

\section{Introduction}

The concept of measurement is at the very heart of quantum mechanics. In
interpretations of quantum mechanics that regard unitary time evolution as
fundamental, measurement is often seen as a problem that needs to be solved. By
contrast, QBism
\cite{Fuchs10a,Fuchs13a,Fuchs14a,Fuchs16a,Fuchs2017,WhereNext,Conversation}
takes measurement as the starting point. According to QBism, a quantum
measurement is an action, any action, that an agent takes on the world external
to herself where the consequences of the action -- the measurement outcomes --
matter to her. A measurement device is conceptually an extension of the agent
who takes the measurement action. The quantum formalism is a tool that any
agent can use to optimize their choice of action.

QBism is an ambitious and ongoing project that aims to discover fundamental
truths about the world \cite{Fuchs16,Merleau} and derive quantum mechanics
from compelling physical principles \cite{Fuchs13a,DeBrota2021}. One of the main
conclusions of QBism is that quantum mechanics is not representational. Even
though its mathematical structure provides deep insights into the nature of
reality, quantum mechanics does not give a description of nature per se, but
functions as a guide to decision making. The quantum formalism does not
describe nature in the absence of agents, but instead is {\it normative}, i.e.,
answers the question of how one {\it should\/} act.

QBist quantum mechanics is rooted in personalist Bayesian probability theory,
first formulated by Ramsey \cite{Ramsey} and de Finetti \cite{deFinetti1989} and
later developed into a full-fledged theory of decision making by Savage
\cite{Savage}. Personalist probability theory provides consistency criteria for
an agent's personal probability assignments that she should strive to satisfy
in order to avoid bad consequences. It is important to recognize that it does
not put any constraints on how one agent's probability assignments should be
related to another agent's probability assignments.

Exactly like personalist probability theory, QBist quantum mechanics is
fundamentally first-person. Compared to probability theory, the quantum
formalism provides additional normative consistency criteria for how the
probabilities an agent assigns to the outcomes of different quantum
measurements should be related \cite{Fuchs13a}. An agent should strive to
satisfy these criteria in their decision making in order to avoid bad
consequences. Similar to Savage's decision theory, QBist quantum mechanics does
not put any constraints on how one agent's assignments of probabilities,
quantum states or measurement operators should be related to another agent's
analogous assignments.

In decision theory, what matters to my decision making are the potential
consequences for myself of my actions. In this spirit, QBism regards
the outcome of a quantum measurement as personal to the agent taking the
measurement action. The outcome of a measurement is thus not something
objective and in principle available to anyone who cares to look. This tenet of
QBism has led to an effortless resolution of recent versions of the Wigner's
Friend paradox \cite{DeBrota2020b}. The tenet implies that, if {\it
  you\/} want to learn the outcome of a measurement {\it I\/} have done, you
will have to ask me what outcome I obtained. Furthermore, because of the
personal nature of both probabilities and measurement outcomes, you and I need
not see the same outcome when we both perform measurements on the same system.

Recently, Khrennikov has argued \cite{khrennikov} that Ozawa's so-called
intersubjectivity theorem \cite{ozawa} invalidates the QBist tenet that a
measurement outcome is personal to the agent performing the measurement. In his
subsequent rebuttal of Khrennikov's claim, Stacey has pointed out \cite{stacey}
that Khrennikov only succeeds in showing that a certain interpretation of
Ozawa's theorem is inconsistent with QBism. Since that interpretation is
incompatible with QBism to start with, Khrennikov's argument fails to challenge
QBism.

Here we build on Stacey's argument to address the question of intersubjective
agreement in QBism. We start in Sec.~2 by contrasting two mathematically
equivalent but conceptually very different ways of describing quantum
measurements. In Sec.~3, we review Ozawa's theorem, show how it can be
interpreted consistently from a QBist perspective, identify the additional
assumptions made by Khrennikov and explain why these are incompatible with
QBism. Section 4 shows that, even though there is never a necessity for two
agents to agree on their respective measurement outcomes, a QBist agent can
strive to create conditions under which she would expect another agent's
reported measurement outcome to agree with hers. The paper concludes with a
brief summary.

\section{General measurements}

The most general $N$-outcome measurement on a $d$-dimensional quantum system
can be described \cite{NielsenChuang} by a set of arbitrary operators ($d\times
d$ matrices) $A_1,\ldots,A_N$ such that
\begin{equation}
  \sum_x A_x^\dagger A_x=1\;. \label{eq:sum}
\end{equation}
The $A_x$ are known as Kraus operators, and the operators $E_x=A_x^\dagger A_x$
form a POVM. The quantum formalism connects the Kraus operators $A_x$, the
state vector $|\psi\rangle$ assigned to the system before the measurement, the
probabilities $p_x$ for the outcomes $x$, and the conditional state vector
$|\psi_x\rangle$ assigned to the system after the measurement
through the formulas
\begin{equation}
  p_x=\| A_x|\psi\rangle\|^2 \label{eq:prob}
\end{equation}
and
\begin{equation}
  |\psi_x\rangle = p_x^{-1/2} A_x |\psi\rangle \label{eq:post} \;.
\end{equation}
This formulation is easily generalized to the case where the agent assigns
density operators rather than pure states to the system. The formulation also
covers measurements of standard quantum observables (von Neumann
measurements): in this case the POVM elements $E_x$ are projectors on the
eigenspaces of the observable, i.e.,
\begin{equation}
  E_x = \Pi_x \;,\;\; \Pi_x\Pi_y=\delta_{xy}\Pi_y \;.
\end{equation}

In QBism, the above constitutes a complete mathematical description of a
quantum measurement. When a QBist agent contemplates a measurement action on a
quantum system, the agent assigns a quantum state to the system and identifies
a set of potential measurement outcomes $x=1,\ldots,N$. The agent then assigns
a Kraus operator $A_x$ to each measurement outcome, with Eq.~(\ref{eq:sum}) as
the only necessary restriction. This step obviously requires intimate knowledge
of the experimental setup and familiarity with the quantum formalism. The
choice of Kraus operators will depend on the agent's prior beliefs, which will
in turn be shaped by the agent's experience as an experimenter. QBism
interprets the conditions~(\ref{eq:prob}) and~(\ref{eq:post}) on the
probabilities and post-measurement states as normative requirements: The agent
should strive to satisfy these conditions to avoid bad consequences. 

It is worth stressing how radically different the above account of quantum
measurement is compared to more mainstream interpretations. In QBism, the Kraus
operators $A_x$ and the POVM elements $E_x$ are not objective states of affairs
there for all to see. They are not ``inherent properties'' of the measurement
device. Instead they are personal to the agent contemplating the
measurement. The same applies to unitaries and Hamiltonians: in QBism, these
are not inherent properties of a physical system but personal to the agent who
is using the quantum formalism to help with her decision making
\cite{Fuchs02}.

Now assume a QBist agent intends to perform the same measurement twice in a
row. By this we mean nothing more than that the agent assigns the same Kraus
operators $A_x$ to both measurements. Let $|\psi\rangle$ be the agent's initial
state. Her probability for obtaining outcome $y$ in the second
measurement, given her outcome was $x$ in the first measurement, is
\begin{equation}
  p_{y|x} = p_x^{-1}\| A_yA_x|\psi\rangle\|^2\;.
    \label{eq:cond}
  \end{equation}
If the $A_x$ are one-dimensional projectors, i.e., if
$A_yA_x=\delta_{yx}A_x$, we have $p_{y|x}=\delta_{yx}$ for all $|\psi\rangle$,
i.e., the agent will expect to get the same outcome in both measurements. But
this is a special case. In general, the agent's probability for getting the
same outcome in both measurements will be less than 1.

Even for projective measurements, i.e., when
$E_x=\Pi_x$ where the $\Pi_x$ are projectors, an agent does not in general
expect a repeated measurement to give the same outcome. The condition
$A_x^\dagger A_x=\Pi_x$ on the Kraus operators is equivalent to $A_x=U_x\Pi_x$
for an arbitrary unitary operator $U_x$, i.e., the Kraus operators need not be
projectors in this case. The condition $E_x=\Pi_x$ by itself places no restrictions on
the conditional probabilities (\ref{eq:cond}).

To summarize, QBism understands quantum measurements from a strictly
single-agent viewpoint. States, Kraus operators, unitaries, probabilities and
measurement outcomes are personal to the agent making the measurement. Even a
single agent does not generally expect to get the same outcome when repeating a
measurement. While another agent contemplating a measurement on the same system
should strive to use the formalism of quantum mechanics consistently, the
formalism itself places no constraints on how the states or Kraus operators of
the two agents should be related. Of course the two agents can communicate and
exchange reports about their respective outcomes. Further below we will
describe a situation where an agent believes that the reported
outcomes will be identical. This changes nothing, however, about the personal
nature of the outcomes.

There is an alternative way of describing a general quantum measurement, which
goes back to Hellwig and Kraus \cite{Kraus1,Kraus2}.
Here, in addition to the quantum system of interest, an
$N$-dimensional system called the meter is introduced, along with a basis
$\{|x\rangle\}$ ($x=1,\ldots,N$). Before the measurement, the joint state of
system and meter is $|\psi\rangle\otimes|1\rangle$. The measurement proceeds in
two steps. First, system and meter are entangled through the action of a joint
unitary (which in QBism expresses the agent's personal judgment),
\begin{equation}
|\psi\rangle\otimes|1\rangle \longrightarrow U(|\psi\rangle\otimes|1\rangle) 
  \;.
\end{equation}
The right-hand side can be expanded in terms of the basis $\{|x\rangle\}$ to
give
\begin{equation}
U(|\psi\rangle\otimes|1\rangle) = \sum_x (A_x|\psi\rangle)\otimes|x\rangle \;,
\end{equation}
where the $A_x$ are linear operators satisfying the normalization
condition (\ref{eq:sum}). Finally a projective measurement is performed on the
meter, which gives outcome $x$ with probability $p_x$ given by
Eq.~(\ref{eq:prob}) and leaves the system in the state
$|\psi_x\rangle=p_x^{-1/2}A_x|\psi\rangle$ as before. 

This way of describing a general quantum measurement is mathematically
equivalent to the description in terms of Kraus operators given above but
conceptually very different. From the QBist perspective, the introduction of the
meter is unnecessary, unmotivated, and confusing \cite{Stacey2023}. It is unnecessary as the
equivalent Kraus-operator formulation does not require the meter. It is
unmotivated as in QBism a measurement device (which the meter is arguably
supposed to model in some way) is simply an extension of the agent. It is
confusing because it suggests that the introduction of the meter somehow
clarifies the measurement process, where in reality it only replaces one
measurement (of the system) by another measurement (of the meter).

\section{Ozawa's theorem}

We will now describe Ozawa's theorem, presented as a purely mathematical
theorem. Following this we will discuss possible interpretations. We present a
slightly simplified version of the theorem, effectively following Khrennikov
\cite{khrennikov}.

Our version of the theorem assumes a tensor product of three finite-dimensional
Hilbert spaces, corresponding to a system of interest and two meters. The
meters are assumed to have fixed initial states $|\xi_1\rangle$ and
$|\xi_2\rangle$. If the initial state of the system is $|\psi\rangle$, a direct
projective measurement on the system will give outcome $x$ ($x=1,\ldots,N$)
with probability
\begin{equation}
  \pi_x=\|\Pi_x|\psi\rangle\|^2\;,
\end{equation}
where the Kraus operators $\Pi_x$ are projectors. Alternatively, one could
first entangle the three systems through a joint unitary $U$ and then perform
measurements on the two meters, with Kraus operators given by projectors
$P^{(1)}_x$ and $P^{(2)}_x$, respectively. In this case, the outcome probabilities
for the meter measurements are
\begin{equation}
  p^{(1)}_x=\|(I\otimes P^{(1)}_x\otimes I)|\Psi\rangle\|^2\;,\;\;
  p^{(2)}_x=\|(I\otimes I\otimes P^{(2)}_x)|\Psi\rangle\|^2 \;,
\end{equation}
where $I$ is the identity and
\begin{equation}
|\Psi\rangle=U\left(|\psi\rangle\otimes|\xi_1\rangle\otimes|\xi_2\rangle\right)
\end{equation}
is the joint state after the unitary interaction. The joint probability for
outcomes $x$ and $y$ for the measurements on the two meters is
\begin{equation}
  p(x,y)=\|(I\otimes P^{(1)}_x\otimes P^{(2)}_y)|\Psi\rangle\|^2\;.
\end{equation}
Notice that the probabilities $p^{(1)}_x$, $p^{(2)}_x$ and $p(x,y)$ refer to the case
that only the two measurements on the meter are carried out.

We now impose the following condition on the otherwise arbitrary unitary $U$,
projectors $\Pi_x$, $P^{(1)}_x$, $P^{(2)}_x$, and initial meter states
$|\xi_1\rangle$, $|\xi_2\rangle$. We assume that, for all initial system states
$|\psi\rangle$, we have
\begin{equation}
  \pi_x=p^{(1)}_x=p^{(2)}_x\;,\;\;x=1,\ldots,N\;.  \label{eq:equalprobs}
\end{equation}  
Ozawa called this condition ``probability reproducibility'' and proved that it
implies that the measurements of the two meters have the same outcome with
probability 1, or
\begin{equation}
  p(x,y) = 0 \;\;\mbox{ unless }\;\; x=y\;. \label{eq:equaloutcomes}
\end{equation}

At first sight this is a surprising result. Why would equality of probabilities
entail equality of outcomes? After all, if you roll two fair dice, most of the
time the outcomes won't match, even though the probabilities for the individual
outcomes are the same. 

The key to understanding Ozawa's result lies in the assumption that equality of
probabilities holds for all initial states $|\psi\rangle$. The essence of the
theorem can be easily grasped by considering the special case that the
projectors $\Pi_x$, $P^{(1)}_x$ and $P^{(2)}_x$ are one-dimensional (the
generalization to arbitrary projectors is straightforward). We can thus write
\begin{equation}\Pi_x=|x\rangle\langle x|\;,\;\;P^{(1)}_x=|\phi^{(1)}_x\rangle\langle
  \phi^{(1)}_x|\;,\;\;P^{(2)}_x=|\phi^{(2)}_x\rangle\langle \phi^{(2)}_x|\;.
\end{equation}

Now let $x$ be any outcome and let the initial system state be $|\psi\rangle =
|x\rangle$.  Then $\pi_x=1$ and thus we must also have that
$p^{(1)}_x=p^{(2)}_x=1$.  It follows that the total state after the unitary
interaction is
\begin{equation}
U\left(|x\rangle\otimes|\xi_1\rangle\otimes|\xi_2\rangle\right)  = 
|\psi_x\rangle\otimes|\phi^{(1)}_x\rangle\otimes|\phi^{(2)}_x\rangle
\end{equation}
where $|\psi_x\rangle$ is a state vector. This holds for any $x$. Hence, for an
arbitrary initial state $|\psi\rangle=\sum_x c_x |x\rangle$, the total
state after the unitary interaction is
\begin{equation}
U\left(|\psi\rangle\otimes|\xi_1\rangle\otimes|\xi_2\rangle\right)  = 
\sum_x c_x |\psi_x\rangle\otimes|\phi^{(1)}_x\rangle\otimes|\phi^{(2)}_x\rangle \;,
\end{equation}
which implies directly that $p(x,y)=0$ unless $x=y$.

As a mathematical result, Ozawa's theorem says nothing about intersubjectivity
or different observers. To arrive at their interpretation, both Ozawa and Khrennikov
have to make the additional assumption that their scenario -- two meters
interacting with a system followed by measurements on the meters -- describes
two different observers measuring the same system observable. We will now show
that it is this additional assumption, not the mathematical theorem, that QBism
is incompatible with.

Our argument is simple. According to QBism, the quantum formalism should be
viewed as a decision-theoretic tool which can be used by any agent. But when an
agent uses the formalism, quantum states, unitaries, measurement operators, and
measurement outcomes are all personal to that agent. Hence, from a QBist
perspective, Ozawa's theorem is about measurements that {\it a single agent},
say, Alice, contemplates performing on a system and two meters. Given the
assumptions of the theorem, Alice expects {\it her\/} outcomes of {\it her\/}
measurements on the two meters to coincide. There is no other agent in the
picture. By construction, QBism is consistent with this interpretation of
Ozawa's theorem. On the other hand, the assumption that the theorem is about
the measurement results of two different observers violates QBism's key tenet
that the quantum formalism should be viewed as a single-agent theory.

The above argument refutes Khrennikov's claim that Ozawa's theorem uncovers an
inconsistency in QBism. It remains silent, however, about what QBism has to say
about intersubjective agreement. We turn to this question in the following section. 

\section{Intersubjective agreement in QBism}

A possible approach to the question of agreement between two agents -- let's
call them again Alice and Bob -- is for Alice to treat Bob as a physical system
and to assign a quantum state to him. There is nothing wrong in principle with
this and, as a thought experiment, it has led to important insights. Examples
include the QBist ``Copernican principle'' \cite{Fuchs13a} and Wigner's friend
\cite{Wigner1961,DeBrota2020b}.

On the other hand, when Alice is not considering thought experiments but uses
the quantum formalism for guidance, e.g., on how to conduct an experiment in a
lab, she will not assign quantum states to other agents. As a matter of fact,
physicists do not normally assign quantum states even to measurement devices
and meters, since assigning a quantum state entails a whole mesh of beliefs
about various measurements that could in principle be performed -- a mesh of
beliefs that physicists almost never hold in a typical experimental situation.

Similarly, physicists rarely if ever have the extraordinarily fine-grained
beliefs entailed by assigning a unitary operator to the interaction between a
system and a measurement device. Since Ozawa interprets the two meters
featuring in his theorem as measurement devices, the scenario on which his
theorem is based must therefore be regarded as a thought experiment. The unitary $U$
assigned to the joint evolution of system and meters is a purely conceptual
object, not an operator that a physicist in a lab would assign and use for
guidance.

So how should Alice, our experimenter in a lab, approach a situation in which
she and her colleague Bob do measurements on the same system?
As we established earlier, once the QBist first-person perspective is adopted,
the quantum formalism does not lead to an automatic agreement between
agents. Instead, agents have to actively create conditions in which they {\it
  expect\/} mutual agreement. Intersubjective agreement is not an automatic
consequence of the quantum formalism, but a goal that agents might strive for.

In practice, Alice would talk to Bob, examine his measurement device, and ask
him what operators, states and probabilities he has assigned. She would check
that his assignments satisfy the constraints of the quantum formalism, take
into account his experimental expertise, and then write down her own (personal)
Bayesian probabilities for the outcomes that she expects Bob to report to
her. Alice would then perform her measurement, wait for Bob to make his
measurement, and finally ask him what outcome he obtained. 

It is important to stress a number of points here. Alice's questions to Bob
should be regarded as actions Alice takes on Bob to elicit answers from
him. These answers do not exist prior to the action; they are created through
the action. Furthermore, whether or not Alice expects Bob's outcomes to agree
with hers is entirely contingent. There are many common scenarios where she
does not expect agreement. For instance, she might come to the conclusion that
Bob's expertise as an experimenter is insufficient.

We will now see how Ozawa's theorem can be modified to provide a simple
condition under which Alice should expect agreement with Bob.  As explained at
the end of Section~2, Ozawa's scenario involving meters that interact unitarily
with the system is, from a QBist perspective, unmotivated, unnecessary, and
confusing. We will instead consider a scenario  where both Alice
and Bob make direct measurements on the same system. We therefore need to adapt
Ozawa's assumption of equal probabilities to our scenario. To make this
concrete, assume that Alice contemplates a measurement on a system and that her
Kraus operators are
\begin{equation}A_x=|\chi_x\rangle\langle\chi_x|\;,\;\; x=1,\ldots,N\;,\;\;
  \langle\chi_x|\chi_y\rangle=\delta_{xy} \;.\
\end{equation}
If her initial system state is $|\psi\rangle$, her probability for outcome $x$
is $|\langle\chi_x|\psi\rangle|^2$, and her post-measurement state, given
outcome $y$, is $|\chi_y\rangle$.

Following Alice's measurement, Bob also performs an $N$-outcome measurement on
the same system. In analogy to the premise of Ozawa's theorem, we assume that,
for any $|\psi\rangle$, Alice's probability for Bob reporting outcome $x$ is
equal to her probability for getting outcome $x$ in her own measurement, i.e.,
is given by
\begin{equation}
  {\rm prob}(\mbox{Bob reports outcome $x$})=|\langle\chi_x|\psi\rangle|^2
  \;. \label{prob:Bobj}
\end{equation}
Now suppose that Alice gets outcome $y$ as a result of her measurement
action. Her post-measurement state is $|\chi_y\rangle$. Thus, her probability
for Bob reporting outcome $y$ following his measurement action is
$|\langle\chi_y|\chi_y\rangle|^2=1$. This means she is subjectively certain that Bob's
report will agree with her measurement outcome.

This simple scenario, which captures the essence of Ozawa's theorem, makes
clear that the assumption (\ref{prob:Bobj}) is the end point rather than the
starting point of Alice's analysis. The probabilities (\ref{prob:Bobj}) may or
may not reflect her degrees of belief. If they do, she will have arrived at
them based on a combination of factors: Bob's record as an experimenter and as
a user of quantum mechanics, her beliefs about Bob's measurement device, Bob's
report to her of his assignments of states and operators, her judgment that he
uses the quantum formalism consistently, and her own assignments of states and
Kraus operators. Once Alice adopts the probabilities (\ref{prob:Bobj}),
consistency requires that she expects Bob's report to agree with her outcome
with probability 1.

\section{Conclusion}

Ozawa's ``intersubjectivity theorem'' is based on a scenario where two meters
are coupled to a system and subsequently measured. In this paper we have
contrasted Ozawa's and Khrennikov's interpretation of this scenario with the
QBist view.  In Ozawa's and Khrennikov's interpretation, the scenario is
assumed to describe two observers measuring the same observable. We have
pointed out that this interpretation does not follow from the mathematical
content of Ozawa's theorem. From a QBist viewpoint, Ozawa's theorem is about a
single agent's expectations regarding the outcomes of that agent's measurement
actions on the system and the two meters. Hence Ozawa's theorem does not
invalidate the QBist tenet that the outcomes of the agent's measurement actions
are personal to the agent. Nothing in the quantum formalism forces two agents
to agree on their respective outcomes. In QBism, intersubjective agreement is
not an automatic consequence of the quantum formalism, but a goal that agents
might strive for. Suitably adapted, Ozawa's theorem can provide an agent with a
condition for when to expect agreement with another agent. 

This work was supported by Grant 62424 from the John Templeton Foundation. The
opinions expressed in this publication are those of the author and do not
necessarily reflect the views of the John Templeton Foundation.

\end{document}